\documentclass{article}

\usepackage{PRIMEarxiv}
\usepackage[utf8]{inputenc} % allow utf-8 input
\usepackage[T1]{fontenc}    % use 8-bit T1 fonts
\usepackage{hyperref}       % hyperlinks
\usepackage{url}            % simple URL typesetting
\usepackage{breakurl}
\usepackage{booktabs}       % professional-quality tables
\usepackage{amsfonts}       % blackboard math symbols
\usepackage{nicefrac}       % compact symbols for 1/2, etc.
\usepackage{microtype}      % microtypography
\usepackage{lipsum}
\usepackage{fancyhdr}       % header
\usepackage{graphicx}       % graphics
\graphicspath{{media/}}     % organize your images and other figures under media/ folder

\title{A data-driven analysis of uk cyber defence
}

\author{
  Justin McKeown \\
  \texttt{Justin McKeown jmckeownuk@gmail.com} \\
}

\begin{document}
\maketitle

\begin{abstract}
Just as computational technologies are a dominant feature of 21st Century life, so too are the various ways in which computers may be exploited for malicious purposes. The ubiquity of computer-based threats is such that, for the UK and other nation states, cyber security is a matter of growing national defence significance. Evidencing this, the UK's most recent integrated review of security, defence and foreign policy mentioned the term  `cyber' 149 times, with the UK prime minister stating therein that by 2030 the UK aspires to be `one of the worlds leading democratic cyber powers'. To fulfill this vision, the UK must first have a sense of the nature of its threat landscape. In order to contribute to this understanding, our research addresses the question: \emph{What are the conditions of the UK’s cyber threat landscape?} In asking this question, we are also seeking to answer a second, more pragmatic question: how can the UK’s cyber defences be improved? There is a tendency when addressing questions of national significance to focus on large complex challenges, such as the threats posed to  the UK by Advanced Persistent Threats (APTs). In this paper we take a different approach, choosing instead to focus on detectable, known and therefore potentially preventable cyber threats, specifically those that are identifiable by the types of malicious scanning activities they exhibit. We have chosen this approach for two reasons. First, as is evidenced herein, the vast majority of cyber threats affecting the lives and business endeavours of UK citizens are identifiable, preventable threats. Thus the potential exists to better improve UK cyber defence by improving how citizens are supported in preventing, detecting and responding to cyber threats; achieving this requires an evidence base. Second, it is potentially useful to build a quantifiable evidence base of the known threat space \textemdash that is to say detectable, identifiable and therefore potentially preventable cyber threats \textemdash to ascertain if this information may also be useful when attempting to detect the emergence of more novel, dangerous cyber threats; for example the kind more readily associated with APTs. Therefore, this research presents an analysis of malicious internet scanning activity collected within the UK between 1st December 2020 and the 30th November 2021. The data was gathered via a custom automated system which collected and processed data from \emph{Greynoise} and \emph{Shodan} APIs, cross referencing it with data from the Office of National Statistics and proprietorial data on UK place names and geolocation. The research was carried out after ethical approval by the University of Oxford's Computer Science Departmental Research Ethics Committee.\\
\\
The data collated during the course of this research has been made available via the Open Science Foundation platform:\\
\hyperlink{https://osf.io/gtbw7/}{https://osf.io/gtbw7/}  
\end{abstract}

% keywords can be removed
\keywords{cyber security \and threat intelligence \and human and societal aspects of security}

\section{Introduction}
Cyber security incidents have become a common feature of 21st Century life. Owing to the ubiquity of computing technologies and the shared nature of cyber space, cyber threats not only affect governments and large corporations, but also SMEs, charities and domestic internet users.  According to the \emph{Cyber Security Breaches Survey 2021} 39\% of UK businesses and 26\% of UK charities reported cyber security breaches in 2021 \cite{GovUK21}.  Further, among medium to large size enterprises, the figure is approximately 65\%, with 51\% of higher income charities experiencing some form of cyber security breach between March 2020 and March 2021 \cite{GovUK21}. In terms of frequency of attack, the report states that 27\% of businesses and 23\% of charities experience an attack at least once per week. It is not only businesses and charities, but also individuals who are affected by cyber security incidents; as an indicator of this the National Cyber Security Centre (NCSC) reported that 2.3 million suspicious emails were forwarded to their reporting service in 2020 \cite{NCSC20}. In considering this report it is evident that beyond the headline grabbing occurrances of cyber attacks on the UK's Parliament \cite{Guardian17, BBC17}, UK universities \cite{Cuthbertson18}, and the mass disruption to the UK's National Health Care System (NHS) as a result of the \emph{WannaCry} malware event \cite{Smart18} UK citizens and organisations face daily challenges in contending with threats in cyber space. Thus defending against cyber threats is a shared challenge, that requires a shared response.\\
\\
In considering how we may make a useful contribution to improving UK cyber defence, we recognise that much scholarly effort is currently being expended in order to develop understanding and approaches to Advanced Persistent Threats (APTs), for example \cite{9868886, 9923774, https://doi.org/10.1002/cpe.6001, ALSARAIREH2022, buzatu_2022}. Arguably, less is being done to formally examine more prosaic cyber threats, the stuff of which makes up the bulk of most citizens and organisations daily cyber challenges. To this end we set out to collect and analyse data concerning detectable, identifiable and therefore potentially preventable cyber threats originating within the UK. In doing this the aim is to develop a body of data detailing the UK's threat landscape and, building on this, develop evidence-based proposals for helping to improve UK cyber defence.\\ 
\\
In considering how we may further focus this research, we recognise there is a difference between targeted and non-targeted attacks. We define targeted attacks as those in which an individual or organisation is wilfully selected as the target for a planned cyber attack on the basis of something more than their systems vulnerabilities. For example an employee who has become disgruntled with their employer may decide to launch a cyber attack against them in recompense for their grievance. We define non-targeted attacks as those in which an individual or organisation is selected as the target for a cyber attack on the basis of the vulnerabilities presented by their accessible systems. For example the \emph{Mirai} malware will scan networks in order to identify suitably configured - and vulnerable - hosts which it can infect. In order to further focus our research, we have decided that we do not want to actively examine targeted attacks and instead we will focus on the non-targeted attacks, which constitute the bulk of cyber threats commonly dealt with by individuals and organisations within the UK.\\
\\ 
A common feature of non-targeted cyber attacks is that they are opportunistic and utilise code in the form of malware to propagate opportunistically across networks and systems. A key technical behaviour of this propagation process is that such malware will perform internet scanning activity in order to identify suitable hosts presenting with the appropriate vulnerabilities. As such focusing on data collection of internet scanning activity originating within the UK will enable us to gain insight into the types and volume of cyber threats within the UK's geographical borders. With this we decided that one year's worth of such data should provide an ample amount of information for this initial study. In collecting this data we initially want to be able to answer simple questions such as how much malicious internet scanning activity is originating from UK based computer systems? What types of malware and other malicious activities are most prevalent and what type of threat might these pose? What types of operating systems are these scanning activities associated with and what regions of the country does this activity originate from?\\
\\
The structure of the remainder of the paper is as follows. We will first provide some further information regarding motivation and  background in order to make evident why we believe this research is necessary and useful. Following from this we provide an outline of our methodology, followed by presentation and analysis of the data we collected between 1st December 2020 and 30th November 2021. Having presented the data we then present questions and implications arising from the data followed, finally, by a conclusion summarising key points in relation to our research questions and and evaluating limits of the study.\\

\section{Motivation and Background}
A significant motivation for the initiation of this research in late 2020 was the findings of our initial scoping analysis of the UK's then current approach to cyber defence. At this time the foremost public document describing the UK's approach to Cyber Security was the \emph{National Cyber Security Strategy 2016-2021} ~\cite{NationalCyberSecurity2016}. Using this document as a starting point we set out to understand how, within the context of cyber, relationships between private citizens, public sector and private sector were envisioned. In so doing we were particularly interested in how responsibility for cyber response was formally and informally distributed across various individuals and organisations within the UK and what events or conditions cause them to become active during a cyber incident of growing intensity. This preparatory analysis surveyed a wide variety of sources including formal models of incident response employed in the UK ~\cite{ISO27001}; UK legislation ~\cite{IntelServicesAct1994, SecServicesAct1989, PoliceReformAct11}; UK government reports and policies ~\cite{GovUK21, HMGovernment21, NationalCyberSecurity2016, NCSCAnnualReview2020}; educational frameworks ~\cite{cybok} and guides ~\cite{CyberEssentials}; as well as frameworks developed by UK professional bodies representing the cyber security industry ~\cite{CiiSecSkillsframework2019, UKSecCouncilSkillsframework2022}. Our analysis indicated that while there was cohesion among stakeholders in specific verticals\footnote{We use the term `Verticals' to refer to business sectors. For example Health care is a vertical. Banking is a vertical. The use of this word in this way is common in business parlance, and is therefore used here.} there appeared to be less cohesion across verticals and thus overall a potential lack of cohesion in the UK's approach to cyber security. Further the key to improving the cohesion in UK cyber security response lay in improving communication and understanding across verticals between technical and non-technical stakeholders. In recognising this, we set out to develop data which could be used to develop a shared understanding of common threats across all verticals, thus mapping the bedrock of the UK's cyber threat landscape. Thus through addressing the previously raised questions regarding types, volumes, location of threats the groundwork could be laid for strategising an evidenced-based approach to more cohesive cyber response.\\  
\\
Since undertaking our initial scoping analysis the UK government has published \emph{The National Cyber Security 2022}, which takes a very different strategic approach than the previous iteration of this strategy. Importantly, the updated version identifies building cohesion across all verticals \textemdash that is to say private citizens, public sector, private sector, government and military \textemdash as a requirement for building better cyber resilience, stating that an `holistic whole-of-society' approach is required to improve UK resilience \cite[p.65]{NationalCyberSecurity2022}. While the updated Cyber security strategy identifies that cohesion must be created it does not detail \emph{how} this should be achieved. On this point we argue that decision making should be predicated on evidence and informed by research. As such the work we have done, despite the refocusing of government cyber policy, remains relevant to the field. Indeed, just as our research is still relevant so too is some of the research that we have examined during the development of our work.  With this in mind, we felt it appropriate to highlight the key research that has informed and further motivated our own research process.\\

\subsection{Defining the problem space: Threat Actors, Target Domains, Defender and Technologies}
One of the biggest challenges we faced when beginning this research was defining a model that adequately represented the cyber problem space. This challenge was not unique to our research and we note that at least early as 2001 researchers were grappling with the challenges of what `cyber space' actually is \cite{10.1111/j.1083-6101.2001.tb00134.x}. While the work of Mitra and Schwartz examined cyberspace as an emerging cultural phenomenon, other researchers examined it specifically from a security perspective. Marrone and Sabatino's \cite{10.2307/resrep28807} analysis of similarities and differences in strategic approaches to cyber space by different NATO member states is significant because it presents an evidence-based argument for the creation of a shared understanding in the form of regulatory and doctrinal framework, that enables better integration of cyber into other defence strategies \cite[p. 3]{10.2307/resrep28807}. In the UK's case this means integration into the vision set out in \emph{The Nation Cyber security strategy 2022}. In considering the work of Marrone and Sabatino, while we agreed with their arguments and analysis, their approach to modelling the cyber defence space was reliant primarily on existing military and policy models and, for our needs, we sought a more simplistic first principles based approach to modeling the problem space. To this end we reduced the problem to threat actors, target domains and defenders.   

\subsection{Threat Actors}
Our initial literature review suggested that the dominant approach to threat actor classification is to define threat actors via their tactics, techniques and procedures (TTPs) and the Indicators of Compromise (IoCs) they leave behind \cite{Noor17-threatActors, Berady21-TTPs}. While this approach is effective from a technical perspective, it is less useful when trying to define threats actors in terms that are useful to both technical and non-technical individuals communicating across differing verticals. To this end, we found the work of Mavroeidis et al. \cite{Mavroeidis21} instructive as it suggested an approach which advocated for the development of commonly agreed-upon controlled vocabularies for characterizing threat actors and their operations. This approach is in dialogue with Marrone and Sabatino's aims of developing a framework that allows for a shared understanding \cite{10.2307/resrep28807} and is a potentially viable strategy for ensuring effective communication across verticals between technical and non-technical stakeholders. Further, examples of this already exist in the intelligence community where `words of estimative probability' \textemdash commonly referred to as WEPs \textemdash are used to confer mathematical probability in plain English \cite{Dhami20}.\\
\\
With the above in mind we searched for an existing shared vocabulary describing threat actors, which is in common use across various verticals and used by technical and non-technical people. To this end we identified the five main categories of threat actor defined by \emph{The National Cyber Security Strategy 2016-2022} as a source of such a taxonomy. These categories are Cyber Criminals, State and State-Sponsored Actors, Terrorists, Hacktivists and Script Kiddies. Cyber criminals are those carrying out two specific types of acts: ‘cyber dependent crimes’ and ‘cyber enabled crimes’ \cite[p. 17]{NationalCyberSecurity2016}. State and state-sponsored threat actors are those threat actors that seek to penetrate UK networks for ‘political, diplomatic, commercial and strategic advantage’ \cite[p. 18]{NationalCyberSecurity2016} at the behest of a nation state. Terrorists are those whose actions fulfil the definition of terrorism given by a range of successive UK legislation passed between 2000 and 2021 \cite[p. 18]{NationalCyberSecurity2016}. Hacktivists are those groups or individuals who are decentralised and issue-orientated and who define targets in response to perceived grievances, ‘introducing a vigilante quality to many of their acts’ \cite[p. 19]{NationalCyberSecurity2016}. Script Kiddies are defined as ‘less skilled individuals who use scripts or programmes developed by others to conduct cyber attacks’ \cite[p. 20]{NationalCyberSecurity2016}. The updated \emph{National Cyber Security Strategy 2022} in no way precludes the use of these terms as part of the UK's current approach to cyber security. The question therefore arises, could these categories offer a starting point for the development of a controlled vocabulary for describing qualities and attributes of threat actors?\\
\\ 
In considering the above further, it is apparent that each threat actor is defined by two key qualities and that these qualities are shared across all five terms. First is the threat actors technological capability. These range from script kiddies who have minimum technical capability, through to nation state actors who have very high technical capability. The second quality is political motivation, which range from cyber criminals who have no political motivation through the personal political motivation of hacktivists to the Political motivation of nation state actors. Thus we note that there may be scope for developing a cyber equivalent to the WEP vocabulary, which utilises technical capability and political motivation as metrics for descriptions of threat actors. The development of such a controlled language would have to be evidence-based and predicated on data which qualified the nature of threat actors. The data we collect during our research could potentially inform the development of such a controlled language and we will therefore return to this subject in the conclusion of this text, once we have presented and analysed our data.\\  

\subsection{Target Domains}
Threat actors do not operate in a vacuum and as such their activities effect people, systems or both. Herein we refer to the impact zones of a threat actors activity as the Target Domain. In reviewing scholarly literature regarding target domains,  we observed that most activity is focused within specific verticals. For example, there is a range of informative literature dealing exclusively with  Industrial control Systems and Critical National Infrastructure \cite{MILLER2021100464, JBAIR2022103611, Samanis9797525, You8472757}. There is decidedly less literature analysing target domains across different verticals. In trying to understand why this is the case, we turned to the field of threat modeling. In reviewing literature in this space \cite{8004867, threatModellingSurvey2014, LEMAY201826, 9821188, 9821458, Shostack15, TATAM2021e05969} it became apparent that because threat actors are commonly defined via TTPs and IoCs, their definition is fundamentally entwined with the targets of their attack. Thus threat actors who attack CNI are identifiable because of the TTPs they exhibit and IoCs they leave when interacting with CNI. Therefore it is logical that most research focused on target domains is domain-specific because the methods used to identify threat actors rely on the specificities of the interaction between specific threat actors and specific target domains.\\ 
\\
In balance we found an alternative approach in the field of Threat Intelligence, and in particular in the work of Mavroeidis and Bormander \cite{Mavroeidis17}. In their research Mavroeidis and Bormander evaluated taxonomies and ontologies within cyber security, ultimately advocating for the development of a multi-layered cyber threat intelligence model. This aligns strongly with the aspirations of our research. In noting this we also note that one of the merits of collecting data regarding malicious scanning activity is that this activity originates from within several different target domains and thus crosses multiple verticals. As such, it creates an evidence base for addressing the question of what would a multi-layered cyber threat intelligence model look like? Further, could it be articulated in a controlled vocabulary that had utility across multiple verticals between technical and non-technical stakeholders? To achieve this we must first identify existing target domains that could be used when analysing our data. In support of this we note that the \emph{National Breaches Survey 2022} \cite{BreachesSurvey22} divides various areas of UK civil endeavour up into the domains of UK micro and small businesses, medium and large businesses, charities, and education. These categories combined with those encompassed by CNI cover a significant amount of potential target domains within UK Society. To these domains we would add that of domestic environments, since this seems overlooked. We will return to this question of target domains, later in the analysis and conclusion of this text.

\subsection{Defenders}
Having considered threat actors and target domains, we also sought to review literature which quantifies the roles, qualities and responsibilities of cyber defenders. We found limited scholarly literature in this domain, and conclude that it is therefore a developing field. Of the scholarly work we identify as relevant, the work of Malviya, Fink et al. was informative \cite{Malviya2011} as it sought to discover qualities that are common to good cyber defenders. In relation to our own study, their work had limited utility primarily because of its focus on military personal. Beyond academia, we found the work of two UK cyber security professional bodies informative. 
The UK Cyber Security Council (UKCSC) \textemdash which the NCSC describe as the ‘voice of the cyber security profession in the UK’ \cite{UKCSC21} \textemdash list 16 specialisms within the field of cyber security \cite{UKCSCSKills22}. In contrast, the Chartered Institute of Information Security `(CIISec) \textemdash which is the other main UK professional cyber security association \textemdash also attempts to quantify the range of specialisms within the UK cyber sector via the publication of a skills framework that lists fourteen different areas of specialisation \cite{CiiSecSkillsframework2019}. In comparing these lists both professional organisations recognise Cyber Security management, Digital Forensics, Threat Intelligence, Secure Systems Architecture, Audit and Assurance, Secure Operations, Incident Response, and Security Testing as categories of cyber security defender endeavour within the UK. In terms of divergence UKCSC recognise Vulnerability Management, Cryptography and Communications Security, Secure Systems Development and Cyber Security Generalist as additional fields of specialism. In contrast CIISec recognise Information Security Governance and Management, Operational Security Management, Data Protection, Privacy and Identity Management, Business Resilience, Information Security Research, and Contributions to the Information Security Professional Development as additional categories of specialism for UK defenders. In considering these areas of convergence and divergence, it is of note that those categories defined by UKCSC are grounded in technical endeavour, whereas those defined by CIISec focus more on categories of management. What is clear from this is that, while there are some areas of agreement among professional bodies, there is no firm consensus on the categories of specialism within UK cyber defence and thus no clearly defined categories of cyber defenders. In nothing this we also note the UKCSC and a number of other bodies are currently working to better define categories of specialism within UK cyber defence. To this end the UK Cyber Body of Knowledge \cite{cybok} \textemdash commonly referred to as \emph{CYBOK} \textemdash has been developed and work is underway to map UK cyber endeavour to this framework. In considering the work of these bodies and the development of CYBOK, the question arises, could analysis of data on types of common threats tackled by UK cyber defenders better inform how we conceive of and define cyber defenders?\\

\subsection{Threat Modelling}
Having defined threat actors, target domains and defenders the natural temptation is to attempt to model the relationship between these things. To this end we could turn to the field of threat modelling and utilise already existing practices, for example \cite{8004867, threatModellingSurvey2014, LEMAY201826, 9821188, 9821458, Shostack15}. However, we will resist the urge to do this and instead turn to the collection, presentation and analysis of our data so as to provide an evidence-base for further addressing the key questions raised thus far in this text. Specifically: through addressing the questions of volume, type and location of malicious scanning activities is it possible to develop an evidence-base upon which one may be able to develop a multi-layered cyber threat intelligence model for the UK, which works across verticals between technical and non-technical stakeholders? Further, could a WEP-style be utilised in this?\\

\section{Methodology}
Having undertaken the above discussed scoping analysis, we designed and built a system that would automate daily collection and analysis of data concerning internet scanning activity originating from within the UK. The system utilised a specially agreed academic API from \emph{Greynoise} \cite{Greynoise22} to enable data collection. IP addresses were extracted from \emph{Greynoise} data and then used to query the \emph{Shodan} \cite{Shodan22} internet search engine. This was done so as to further enrich the data gathered via \emph{Greynoise}. Information from both \emph{Greynoise} and \emph{Shodan} was then further contextualised using data supplied by the \emph{Office of National Statistics} (ONS) and proprietary geolocation data purchased from the UK company \emph{TownList} \cite{Townlist22}. This data collection and contextualising process took place on a daily basis. Each day at 10:07AM the process would begin with an automated request to \emph{Greynoise} for the entirety of their previous days data concerning internet scanning activity originating from within the UK. After running an initial testing of the system for one month, we began formal data collection on the 1st December 2020 and ceased formal collection the 30th November 2021. Through analysis of the data and by examining it in light of the material covered in our initial scoping analysis we were able to develop a series of insights and recommendations regarding our reserarch questions.\\

\section{Data presentation and analysis}
In this section we present the data gathered along with supplemental information to aid with its interpretation. Though before doing so it is important to discuss the quality of the data being analysed.\\

\subsection{Data Quality}
Over the 365 days of data collection, \emph{Greynoise} reported that it detected and tagged a total of 851,972 internet scanning activities originating within the UK. From \emph{Greynoise's} 851,972 detection records we were able to retrieve 491,668. The missing data is attributed to \emph{Greynoise} capping the number of results it returned via the API it had made available for our research purposes. For clarity, of the missing 360,304 records, 301,467 are associated with unprecedented swells in internet scanning activity on the 18th and 19th of December 2020 and on the 15th and 16th April 2021. It is of note that these dates correlate with the \emph{SolarWinds} and \emph{Microsoft Exchange Server} attacks. These events may well be the reason for the unprecedented swell in internet scanning activity on these dates. While the \emph{Greynoise} API did not allow us to capture complete datasets for these dates we did manage to retrieve partial datasets for each, consisting of 10,000 entries per day. The remaining 58,837 unaccounted for records are spread throughout the period from 4th July 2021 to 30th November 2021. Aside from the omissions noted, the data collected between 1st December 2020 and 3rd July 2021 is complete.\\

\subsection{Identifying UK based scanning activity}
Of the 491,668 complete records obtained, \emph{Greynoise} classified 211,460 records as being from ‘unspoofable’ IP addresses. \emph{Greynoise} classifies an IP address as ‘unspoofable’ when it has observed a complete TCP authentication process. Thus, we can conclude that 211,460 of the records account for activity originating from within the UK, while the remaining 280,208 may or may not originate from a UK based IP address. In considering this, for the purposes of accuracy, we will continue to focus only on activity which originates from ‘unspoofable’ UK IP addresses.\\

\subsection{Malicious, Benign and unclassifiable scanning activity}
Of the 211,460 ‘unspoofable’ records, \emph{Greynoise} classified 7,410 as `benign'; 77,677 as `malicious' and the remaining 126,373 as `unknown'. \emph{Greynoise} classifies activity as benign when it recognises the source of the scanning activity; for example, search engines such as Google. Where an entity exhibits behaviour that explicitly evidences malevolent intent it is classified as malicious; for example attempted use of the \emph{Bluekeep} exploit \cite{Bluekeep} on one of \emph{Greynoise's} cloud-based sensors. When scanning activity falls into neither category it is classified as ‘unknown’. Therefore some of the data classified as `unknown' may in fact be malicious activity that, at the time of collection, could not be accurately verified as such. The inverse of this statement is also true. Table 1 presents a breakdown of the monthly volume of internet scanning activity originating from the UK along with a breakdown of their \emph{Greynoise} classification. In examining this data it is notable that the volume of `malicious' and `unknown' scanning activity is significantly larger than that of `benign' scanning activity.\\ 
\\

\begin{table*}
  \label{tab:commands}
  \begin{tabular}{cccccl}
    \toprule
    Month & Benign & Malicious & Unknown & Total\\
    \midrule
    Dec-20 &	536 &	5,872 &	12,343	& 18,751\\
    Jan-21 &	699 &	7,180 &	16,339	& 24,218\\
    Feb-21 &	1,289 &	5,633 &	8,272	& 15,194\\
    Mar-21 &	640 &	7,135 &	14,444	& 22,219\\
    Apr-21 &	803 &	6,916 &	14,647	& 22,366\\
    May-21 &	816 &	7,569 &	13,145	& 21,530\\
    Jun-21 &	606 &	7,117 &	9,305	& 17,028\\
    Jul-21 &	611 &	6,959 &	7,347	& 14,917\\
    Aug-21 &	316 &	5,249 &	5,814	& 11,379\\
    Sep-21 &	622 &	6,357 &	6,705	& 13,684\\
    Oct-21 &	220 &	6,368 &	9,015	& 15,603\\
    Nov-21 &	252 &	5,322 &	8,997	& 14,571\\
    \bottomrule
  \end{tabular}
   \caption{Monthly volume of internet scanning activity originating within the UK classified by Greynoise as benign, malicious and unknown. Months marked with * have incomplete datasets}
\end{table*}

\subsection{Types of malicious scanning activity}
For the purposes of transparency, \emph{Greynoise} also makes a number of distinctions in its tagging process which would be highly useful to a SOC analyst, but which are overly granular for the purposes of our analysis. For example \emph{Greynoise} provides separate tags for \emph{Mirai} malware and some of its derivatives. Further \emph{Greynoise} also distinguishes between \emph{Eternalblue} and activity which is highly probably \emph{Eternalblue}. We have therefore chosen to ignore these minor distinctions and combine the data for these tags. This combination gives a detection count for \emph{Mirai} of 27,361 and 18,107 for \emph{Eternalblue}. With all of the above interpretations of the original \emph{Greynoise} data accounted for, it is significant that of the 279 unique subcategories of scanning activities, The top 10 of these account for 61.85\% of all malicious scanning activity originating from UK IP addresses during the period of study. Of these top 10, the top 5 are responsible for 44.71\% of all malicious scanning activity originating within the UK (see Table 2). In making these claims it is important to be clear that they are based on counting the number of times various scanning activities were tagged by \emph{Greynoise} and thus the claim is limited by the fact that the data is collected from a single source.  For thoroughness, below we provide a brief summary of what the top 10 detected malicious scanning activities are and the threats they represent.\\

\begin{table*}
  \label{tab:commands}
  \begin{tabular}{ccccl}
    \toprule
    Rank & Malicious Scanning Activity & Total Times Tagged & Percentage of Total Detections\\
    \midrule
    1	& Mirai	                            & 27,259    & 11.98\%\\
    2	& Web Crawler	                    & 23,284    & 10.24\%\\
    3	& Telnet Bruteforcer	            & 18,338    & 8.06\%\\
    4	& Eternalblue	                    & 18,107    & 7.96\%\\
    5	& Generic IoT Brute Force Attempt	& 14,707    & 6.47\%\\
    6	& SSH Worm	                        & 10,545    & 4.64\%\\
    7	& SMBv1 Crawler	                    & 9,136	    & 4.02\%\\
    8	& SSH Bruteforcer	                & 8,359	    & 3.68\%\\
    9	& Telnet Scanner	                & 6,400	    & 2.81\%\\
    10	& HTTP Alt Scanner	                & 4,530	    & 1.99\%\\
    \bottomrule
  \end{tabular}
   \caption{By volume of Greynoise observations, the top ten most commonly detected malicious scanning activities detected within the UK within the Period Dec '20 to Dec '21. These account for 61.85\% of all malicious scanning activity originating within the UK.}
\end{table*}
\subsubsection{Mirai}
By volume of activity, \emph{Mirai} malware and its variants are the most common source of malicious scanning activity detected within the UK during the research period. \emph{Mirai} commonly infects so-called ‘smart’ devices that run ARC processors \cite{MiraiBotnet}. When \emph{Mirai} infects a device, it turns it into a node in a botnet network and later uses it to launch DDoS attacks. \emph{Mirai} spreads by scanning the internet looking for devices capable of hosting it. If the default name and password is not changed on such devices, then \emph{Mirai} is able to log into the device and infect it. \emph{Greynoise} applies the \emph{Mirai} classification to devices that have exhibited behaviour indicative of infection with \emph{Mirai} or a Mirai variant \cite{GreynoiseTags22}. The risk arising from the \emph{Mirai} threat is the hijacking and mobilisation of UK-based computer devices to launch attacks both within and outside the UK.\\

\subsubsection{Web Crawler}
The term \emph{web crawler} is somewhat ambiguous when used in the context of malicious scanning activity. A web crawler crawls the web indexing sites using criteria set by the crawler’s authors. Given the lack of further detail available on the types of web crawling behaviours observed by \emph{Greynoise}, we can only assume that the activity is indicative of intelligence gathering and the reconnaissance part of the cyber kill chain. The risk arising from these threats is therefore the targeting of individuals, organisations or websites via information gathered through this profiling activity. \\

\subsubsection{Telnet Bruteforcer and Telnet Scanner}
Telnet scanners and Telnet bruteforcers both target the Telnet application protocol, though in slightly different ways. Both search the internet for servers exposing the Telnet to the web. When a server is discovered, the Telnet scanner will attempt to fingerprint any services running; a typical example of this is afforded via the \emph{Metasploit’s} Telnet auxiliary modules \cite{TelnetScanner}. Going beyond fingerprinting, the Telnet brute forcer will attempt to brute force the target system’s password.  Telnet scanning activity may well be a precursor to a Telnet bruteforce attempt. Further, both activities are highly likely to be a precursor to other forms of malicious activity. The risk therefore is further exploitation of the server, undermining the confidentiality, integrity  and availability of services hosted by the compromised system. The compromised system may also be used as a platform from which to launch more sophisticated attacks, such as those discussed above, against other servers.\\

\subsubsection{Generic IoT Brute Force attempt}
\emph{Greynoise} applies the tag \emph{generic IoT brute force attempt} to any IP that has been observed attempting to brute force IoT devices via Telnet or SSH with generic default credentials \cite{GreynoiseTags22}. Such activity therefore bears relationship to both \emph{Mirai} and Telnet scanners both in terms of threat and behaviour; and accordingly, in terms of risk.\\

\subsubsection{Eternalblue and SMbv1 Crawlers}
\emph{Eternalblue} is the name given to an exploit that targets a vulnerability (MS17-010) in Microsoft’s Sever Message Block (SMB) protocol. This exploit was a feature of both \emph{WannaCry} and \emph{NotPetya} \cite{Burges17}. The exploit was originally developed by America's National Security Agency (NSA), before being stolen and leaked to the wider internet by the \emph{Shadow Brokers} hacking group. \emph{EternalBlue} targets multiple versions of Windows and enables attackers to gain remote access to infected devices. \emph{Greynoise} applies this classification to IP addresses observed trying to infect devices across the internet using the EquationGroup’s \emph{EternalBlue} exploit \cite{GreynoiseTags22}. Related to this, \emph{Greynoise} classifies \emph{SMBv1 crawlers} as those IP addresses which have been observed crawling the internet for SMBv1. Thus, the latter could in some circumstances be an activity that prefigures the former.\\
\\
The threat posed by compromise of the SMB protocol – via \emph{EternalBlue} or other means - is multifaceted and dependent on the accordances of the infected device, and what other devices it may be connected to. Worryingly, this could include Critical National Infrastructure as was the case with \emph{NotPetya}. The risk therefore is compromise and disruption of services across the entire spectrum of domains in which computing devices are used.\\  

\subsubsection{SSH Worm and SSH Brute Forcer}
\emph{Greynoise} applies the tag \emph{SSH worm} to IP addresses that have been observed attempting to brute force SSH server credentials and that also exhibit signs indicating that it was compromised and is operating on behalf of a another actor \cite{GreynoiseTags22}. The \emph{SSH Brute Forcer} classification tag is only applied to devices observed attempting to brute force SSH credentials. The risks arising from SSH worm activity and SSH brute forcing are the compromise of the confidentiality, integrity, or availability of SSH worm compromised servers and their further exploitation via malicious actors. This may include more severe attacks on other systems, organisations or even individuals depending on the nature of services and data hosted on the compromised machine.\\

\subsubsection{HTTP Alt Scanner}
This tag is applied to entities observed scanning web servers for open alternative http ports. This may be as part of an attempt to compromise servers hosting web services through non-standard ports. If a server is compromised via such a service then it may be used as a platform from which the attacker may launch other, more sophisticated attacks.\\
\\

\section{Scanning by Operating System}
Having considered the various forms of malicious scanning activity, it is also insightful to consider the types of operating system most frequently associated with each top ten scanning activity. We present this  data in Table 3. In examining this information it is evident that computers running Linux Kernel 2.2-3.x are most frequently the source of malicious scanning activity. Linux Kernel 2.2 dates from 2000 with the earliest version of Linux Kernel 3 dating from 2012. The last version of Linux Kernel 3.x was released in 2015. This therefore suggests that the computers running these operating systems have not been updated due to neglect, or else they cannot be updated owing to either software dependencies or the critical nature of roles they perform while running.\\

\begin{table*}
  \label{tab:commands}
  \begin{tabular}{ccccl}
    \toprule
    Top Ten Rank & Malicious Scanning Activity & Most Tagged Operating System & Total Times Tagged\\
    \midrule
        1   & Mirai	                            & Linux 2.2-3.x	    & 12,291\\
        2   & Eternalblue     	                & Windows 7/8	    & 11,241\\
        3   & Web Crawler	                    & Linux 2.2-3.x	    & 10,298\\
        4   & Telnet Bruteforcer	            & Linux 2.2-3.x	    & 7,412\\
        5   & SMBv1 Crawler	                    & Windows 7/8	    & 7,091\\
        6   & Generic IoT Brute Force Attempt	& Unknown	        & 6,806\\
        7   & SSH Worm	                        & Linux 2.2-3.x	    & 5,044\\
        8   & SSH Bruteforcer	                & Linux 2.2-3.x	    & 4,046\\
        9   & Telnet Scanner	                & Linux 2.2-3.x	    & 3,515\\
        10  & HTTP Alt Scanner	                & Linux 2.2-3.x	    & 2,005\\
    \bottomrule
  \end{tabular}
   \caption{Operating systems most frequently tagged against each top ten scanning activity within the UK for the Period Dec '20 to Dec '21.}
\end{table*}

Beyond the top ten detected malicious scanning activities, it is instructive to examine the breadth of operating systems from which malicious scanning activity appears to originate. Within our study we detected 14 different variants of operating system associated with malicious scanning activity. There were also a multitude of scanning activities originating from operating systems which we could not resolve to a specific known fingerprint. We have categorised data originating from the later as being attached to `unknown' operating systems. In doing this we note that these unknown operating systems may be one of those already listed in the table, it is simply the case that the operating system could not be resolved at the time of detection. We present this data in Table 4.\\
\\
In considering the implications of the data listed in Table 4, it is important to  note that the predominance of Linux in the listings does not imply that it is a less secure operating system. Indeed, FreeBSD is a Linux operating system and ranks below that of Mac OS X for times tagged in relation to malicious scanning activity.  The frequency with which an operating system is associated with malicious scanning activity is most likely attributable to a combination of factors. Firstly the proliferation of the operating system and secondly the context in which it ids used. For example, Linux is often the operating system of choice for server administrators setting up services which can be accessed from the internet.

\begin{table*}
  \label{tab:commands}
  \begin{tabular}{ccccl}
    \toprule
    Rank & Operating System & Total Times Tagged & Total Times Tagged in\\
    & & & relation to a Top Ten Scanning Activity\\
    \midrule
        1 & Linux 2.2-3.x				& 90,991	& 49,841\\
        2 & unknown						& 32,023	& 28,317\\
        3 & Windows 7/8					& 33,632	& 18,885\\
        4 & Linux 3.1-3.10				& 15,610	& 12,273\\
        5 & Linux 3.11+					& 27,637	& 11,379\\
        6 & Linux 2.2.x-3.x (Embedded)	& 8,662		& 5,048\\
        7 & Linux 3.x					& 3,691		& 2,919\\
        8 & Linux 2.4.x					& 2,649		& 2,342\\
        9 & Windows XP					& 5,225		& 2,000\\
        10 & Windows 2000				& 5,591		& 1,049\\
        11 & Linux 2.4-2.6				& 712		& 411\\
        12 & Linux 2.6					& 502		& 347\\
        13 & Windows NT kernel 6.x		& 293		& 192\\
        14 & Mac OS X					& 220		& 46\\
        15 & FreeBSD					&	4		& 0\\
    \bottomrule
  \end{tabular}
   \caption{Operating systems most frequently tagged against each top ten scanning activity within the UK for the Period Dec '20 to Dec '21.}
\end{table*}

%%Unique IP Addresses
\section{Volume of unique IP addresses scanning}
Within the period of study a total of 18,799 unique IP Addresses were detected performing malicious scanning activity  from within the UK. Almost all of these unique IP addresses were associated with multiple types of malicious scanning activity. Table 3 presents the number of the unique IP addresses associated with each of the top ten scanning activities. In reading between Tables 2 and 5 it is evident that, for example, 8,017 IP addresses were associated with 27,259 detected \emph{Mirari} scanning activities.\\

\begin{table*}
  \label{tab:commands}
  \begin{tabular}{cccl}
    \toprule
    Malicious Scanning Activity & Number of Unique IPs & Percentage of total Unique IPs\\
    \midrule
    Mirai	                            & 8,017 &	42.71\%\\
    Web Crawler	                        & 5,878 &	31.31\%\\
    Telnet Bruteforcer	                & 4,508 &	24.01\%\\
    Generic IoT Brute Force Attempt	    & 3,297 &   17.56\%\\
    SSH Bruteforcer	                    & 2,388 &	12.72\%\\
    SSH Worm	                        & 2,021 &	10.77\%\\
    Telnet Scanner	                    & 1,848 &	9.84\%\\
    Eternalblue	                        & 979   &	5.22\%\\
    SMBv1 Crawler	                    & 977   &	5.2\%\\
    HTTP Alt Scanner	                & 955   &	5.09\%\\
    \bottomrule
  \end{tabular}
   \caption{the volume of Unique IP addresses associated with the top ten most commonly detected malicious scanning activities within the UK for the Period Dec '20 to Dec '21.}
\end{table*}

\subsection{IP Addresses data interpretation}
In considering the data presented in Table 5 it is important to note for less technical readers that unique IP detections do not neatly translate into the number of unique machines performing the associated activity. To be explicit: just because 8,017 unique IP addresses were detected performing behaviour indicative of \emph{Mirai} malware scanning activity does not mean that there are 8,017 computers infected with \emph{Mirai} scanning the internet looking for new hosts to infect. The reason IP address count does not neatly translate to the number of machines performing a given activity is because of the way in which IPv4 addresses are utilised as a finite resource for addressing networks and systems across the global internet. It is important to clarify how IPs are used as a resource so that we are precise in defining what the collected data can and cannot tell us.\\

\subsection{ASNs, IP Pools and ISPs}
An Autonomous System Number (ASN) is an identification number given to the controller of a routable IP address range managed by a single entity or organisation. This organisation is responsible for the management, supervision and control of activities within the given IP range. An Internet Service Provider (ISP) provides one such instance of such an organisation. Thus, for example, the controller of ASN  AS9105 is UK-based ISP \emph{TalkTalk Communication Limited} who are responsible for 3,032,320 IPs across a number of address ranges \cite{TalkTalkASNDetails}.\\
\\
In terms of day to day management, the IP addresses in a given ASN are essentially a pool of resources that are assigned and unassigned to different consumers for varying periods of time. For example, the IP addresses allocated to domestic internet users are frequently renewed by their ISPs. Thus a domestic customer's external network IP address may change every few days. Conversely, some business users may require fixed IP addresses to enable delivery of their services and thus may operate from fixed IP addresses all year round. When a customer leases an IP from an ASN controller --- most commonly via and ISP --- they may in turn use it to create a subnet. Thus a single IP address in an ASN may be the gateway address of a network comprised of hundred of different computing devices. The processes of assigning and unassigning IPs combined with the ability of those leasing IPs to subnet them is the reason  why we cannot neatly equate one IP address to one machine when reading the collected data.\\ 
\\
With the above in mind, while it is not possible within the scope of the study to neatly equate the number of observed IP addresses to the number of machines performing malicious scanning activity, it is theoretically possible to use the collected data to not only count but also locate the machines performing the observed malicious scanning activity. Doing this would require retrieving additional data from multiple ASN controllers, though it is possible because as part of the management of their networks, ASN controllers are required to keep records of which IP addresses are assigned to which legal entities (i.e. people, businesses) including the date and time these were assigned and unassigned. As such, through direct working with ASN controllers, it would be possible to calculate the number of IP addresses responsible for each scanning activity and indeed identify the legal entities associated with these IP addresses; one would simply need to be important enough for multiple ASN controllers to be willing to collaborate on such a project. For example, one may need to be a government agency.\\
\\

\subsection{Malicious Scanning by ASN Controller}
While the above calculations are beyond the scope of the data available to us, it it is possible to use the collected data to calculate the volume of scanning activity associated with various ASN controllers networks. We can do this by taking the IP addresses associated with malicious scanning activity and grouping these by ASN controller. In turn we can then use the existing mapping of IPs to malicious scanning activity to map malicious scanning back to ASNs. If we examine this data on a monthly basis we are able to gain insight into the number of times particular types of malicious scanning activity were tagged within each ASN controllers IP pool. From this we infer the frequency with which various activities were detected.\\
\\
Within the data collected we identified 2,298 unique ASN controllers. Of these 2,298 ASN controllers, malicious scanning activity was observed to originate from just 406 of them. This suggests that 100\% of the activity discussed herein are originating from IPs governed by 17.7\% of ASNs.  Table 6 provides a summary of the 10 ASN Controllers in which malicious scanning activity was tagged most frequently. In contrast Table 7 provides a summary of the 10 ASN controllers in which the top ten malicious scanning activities were tagged in most frequently. It is of note that 9 of the providers feature in both Tables 6 and 7.\\ 
\\
It is important to note that just because malicious scanning activity has not been observed to originate from an ASN controllers network does not mean that it is free of malicious scanning activity. Rather, it could equally be true that the type of malicious scanning activity originating in the network was not classifiable as malicious by \emph{Greynoise} and therefore has evaded consideration in our study. Thus it may simply be that the nature of the nature of the activities conducted on the ASN's listed in Tables 4 and 5 is conducive to the types of infections we are examining. A further consideration is that, because our data is drawn from a single source \textemdash \emph{Greynoise} \textemdash, then the concentration of scanning activity in a small percentage of ASNs may be reflective of the locations of \emph{Greynoise's} detection systems. Clarifying \emph{exactly} why we see this imbalance across ASNs is beyond the scope of this study, though we note that is worthy of further investigation in future research.\\ 
\\

\begin{table*}
  \label{tab:commands}
  \begin{tabular}{cccccl}
    \toprule
    ASN Owner 	& ASN 		& Times malicious 	& Unique IPs 	& Number of IPs\\
		& Number 	& scanning tagged 	&		& in ASN pool\\
    \midrule
    DigitalOcean, LLC				   & 	AS14061		&	30,477		&	3,162	&	2,724,352\\
    TalkTalk Communications Limited		& 	AS13285		&	30,412		&	4,765	&	3,521,792\\
    British Telecommunications PLC		& 	AS2856		&	25,579		&	3,675	&	11,903,232\\
    Virgin Media Limited				& 	AS5089		&	20,620		&	1,936	&	9,470,976\\
    Amazon.com, Inc.					& 	AS16509		&	12,934		&	941		&	43,000,320\\
    KCOM GROUP LIMITED					& 	AS12390		&	11,633		&	1,048	&	262,656\\
    Sky UK Limited						& 	AS5607		&	10,498		&	1,872	&	7,859,968\\
    Alibaba (US) Technology Co., Ltd.	& 	AS45102		&	7,116		&	502		&	2,479,360\\
    OVH SAS								& 	AS16276		&	6,890		&	691		&	4,091,904\\
    Plusnet								& 	AS6871		&	6,686		&	569		&	2,138,368\\
    \bottomrule
  \end{tabular}
   \caption{ASN Controllers ranked by volume of Number of times malicious scanning activity was tagged in their IP pool within the study period Dec '20 to Dec '21.}
\end{table*}

\begin{table*}
  \label{tab:commands}
  \begin{tabular}{cccccl}
    \toprule
   ASN Owner 	& ASN 		& Times malicious 	& Unique IPs 	& Number of IPs\\
		& Number 	& scanning tagged 	&		& in ASN pool\\
    \midrule
    TalkTalk Communications Limited		& 	AS13285		& 	27,386	& 	4,526	& 	3,521,792\\
    British Telecommunications PLC		& 	AS2856		& 	19,962	& 	3,059	& 	11,903,232\\
    Virgin Media Limited				& 	AS5089		& 	13,613	& 	1,662	& 	9,470,976\\
    DigitalOcean, LLC					& 	AS14061		& 	11,295	& 	2,940	& 	2,724,352\\
    KCOM GROUP LIMITED					& 	AS12390		& 	10,850	& 	1,027	& 	262,656\\
    Sky UK Limited						& 	AS5607		& 	7,643	& 	1,749	& 	7,859,968\\
    Amazon.com, Inc.					& 	AS16509		& 	4,761	& 	876		& 	43,000,320\\
    Plusnet								& 	AS6871		& 	4,630	& 	499		& 	2,138,368\\
    OVH SAS								& 	AS16276		& 	3,510	& 	584		& 	4,091,904\\
    Daisy Communications Ltd			& 	AS5413		& 	2,230	& 	109		& 	743,680\\
    \bottomrule
  \end{tabular}
   \caption{ASN Controllers ranked by volume of Number of times top ten malicious scanning activity was tagged in their IP pool within the study period Dec '20 to Dec '21.}
\end{table*}

\section{Scanning by geographic location}
Having examined types of malicious scanning activity, volume of associated IP addresses and the ASN 	controllers associated with these IPs, it is interesting to also consider the geographic locations to which these IPs belong. In doing this, it must be noted that our ability to resolve IPs to geophysical locations is largely reliant on \emph{Greynoise's} tagging of the city location, region and country from which various scanning activities originate. Table 8 presents a summary of the top cities in the UK from which malicious scanning activity appears to originate. Table 9 presents similar data, but with a focus on the top 10 scanning activities introduced in Table 2.\\ 	

\begin{table*}
  \label{tab:commands}
  \begin{tabular}{cccl}
    \toprule
   Rank & Location & Total times Malicious Scanning Tagged in location\\
    \midrule
1		& London-England			& 111,494\\
2		& Manchester-England		& 6,443\\
3		& Birmingham-England		& 4,421\\
4		& Bexley-England			& 4,312\\
5		& Cardiff-Wales				& 3,031\\
6		& Leeds-England				& 2,641\\
7		& Liverpool-England			& 2,344\\
8		& Glasgow-Scotland			& 2,026\\
9		& Norwich-England			& 1,734\\
10		& Salford-England			& 1,622\\
    \bottomrule
  \end{tabular}
   \caption{Top 10 UK Locations ranked by volume of Malicious Scanning activity originating from them within the study period Dec '20 to Dec '21.}
\end{table*}

In examining the data presented in Tables 8 and 9 it is apparent that major UK cities are the main sites from which much malicious scanning activity appears to originate. Upon analysis of Office for National Statistics (ONS) data, population does not seem to be a defining factor in the frequency with which malicious scanning is tagged in various UK locations. It is more probable that population size and the infrastructure of internet routing plays a contributing roles with regards the frequency of tagging in relation to geographic locations. However, testing this proposal is beyond the scope of what has been possible so far with this research.\\

\begin{table*}
  \label{tab:commands}
  \begin{tabular}{cccl}
    \toprule
   Rank & Location & Total times Malicious Scanning Tagged in location\\
    \midrule
1		& London-England			& 54,127\\
2		& Manchester-England		& 3,342\\
3		& Birmingham-England		& 2,996\\
4		& Cardiff-Wales				& 2,269\\
5		& Bexley-England			& 2,056\\
6		& Leeds-England				& 1,837\\
7		& Liverpool-England			& 1,696\\
8		& Southampton-England		& 1,380\\
9		& Norwich-England			& 1,304\\
10		& Luton-England				& 1,193\\
    \bottomrule
  \end{tabular}
   \caption{Top 10 UK Locations ranked by volume of Top Ten Malicious Scanning activity originating from them, within the study period Dec '20 to Dec '21.}
\end{table*}

\section{Scanning by Police Region}
In further considering the geographic spread of malicious scanning activity it is interesting to note that Section 77 of the \emph{Police Reform and Social Responsibility Act (2001)} \cite{PoliceReformAct2011} outlines seven national strategic policing requirements, which the Home Secretary requires regional police forces to prioritise. Among these are serious organised crime, child sexual exploitation and national cyber security incidents \cite{PolicingInTheUK}. In light of this increasing demand on police to develop their capcity to cyber crime it is therefore insightful to consider how the data examined so far maps into UK policing regions. Table 8 presents this information for the entirety of the UK's policing services.\\
\\
In examining the data in Table 10 in relation to that in Table 8 and 9, it is of note that there is some approximate mapping between the ranking of large cities by observed malicious scanning activity and the policing regions within which they are located. For example, City of London police and Metropolitan  police have jurisdiction over London and greater London. Thus, both London and Bexley which are cited in Tables 8 and 9, fall within their jurisdiction. Similarly, Birmingham is covered by West Midlands police, Cardiff by South Wales police,  Leeds by West Yorkshire police, Liverpool by Merseyside police and Southampton by Hampshire police. Norwich lies within the jurisdiction of Norfolk police, which is ranked 15th. Only the ranking of Bedfordshire police in 30th position, who have jurisdiction over Luton, seems anomalous.\\
\\
In examining this data it is evident that, overall, City of London is the policing region from which by far the most malicious scanning activity was seen to originate from. The reasons for this are unclear, but the statistics are notable and prompt future study.

\begin{table*}
  \label{tab:commands}
  \begin{tabular}{ccccccl}
    \toprule
   Rank 	& Policing 	& Total Malicious 	& Total 		& Total Top ten 		& Total unique IPs\\
			& Region 	& Scanning 		& unique IPs 		& malicious scanning 	& assoc with top ten scanning\\
    \midrule
    1   &     London, City of	    & 112,103    &	8,262    &      54,300   &      7,033\\
    2   &     Greater Manchester	& 10,640     &	806     &  	    6,145    &      721\\
    3   &     Metropolitan Police	& 10,489     &	1,070    &      5,709    &      947\\
    4   &     West Midlands	        & 8,138      &	859     &  	    5,534    &      725\\
    5   &     Police Scotland	    & 6,898      &	795     &  	    4,443    &      752\\
    6   &     West Yorkshire	    & 4,885      &	475     &  	    3,468    &      417\\
    7   &     Hampshire	            & 4,143      &	374     &  	    3,057    &      348\\
    8   &     South Wales	        & 3,896      &	441     &  	    2,837    &      375\\
    9   &     Thames Valley	        & 3,859      &	334     &  	    2,796    &      294\\
    10  &    Devon \& Cornwall	    & 3,374      &	393     &  	    2,615    &      332\\
    11  &    Merseyside	            & 3,365      &	458     &  	    2,487    &      424\\
    12  &    PSNI	                & 3,195      &	280     &  	    2,172    &      272\\
    13  &    Avon and Somerset	    & 3,160      &	318     &  	    2,279    &      257\\
    14  &    Norfolk	            & 2,955      &	327     &  	    2,323    &      307\\
    15  &    Nottinghamshire        & 2,801      &	286     &  	    1,899    &      249\\
    16  &    Kent	                & 2,617      &	242     &  	    2,015    &      226\\
    17  &    South Yorkshire        & 2,513      &	332     &  	    1,786    &      290\\
    18  &    Essex	                & 2,472      &	263     &  	    1,906    &      248\\
    19  &    Dorset	                & 2,432      &	189     &  	    1,732    &      177\\
    20  &    Surrey	                & 2,270      &	115     &  	    1,934    &      111\\
    21  &    Lancashire	            & 2,127      &	252     &  	    1,722    &      236\\
    22  &    Lincolnshire	        & 2,090      &	234     &  	    1,560    &      220\\
    23  &    Staffordshire	        & 1,890      &	202     &  	    1,515    &      190\\
    24  &    Northumbria	        & 1,889      &	283     &  	    1,387    &      264\\
    25  &    Leicestershire	        & 1,852      &	141     &  	    1,474    &      121\\
    26  &    Sussex	                & 1,810      &	217     &  	    1,466    &      209\\
    27  &    Wiltshire	            & 1,770      &	120     &  	    1,394    &      112\\
    28  &    Northamptonshire	    & 1,550      &	112     &  	    1,039    &      105\\
    29  &    West Mercia	        & 1,493      &	190     &  	    973      &       170\\
    30  &    Bedfordshire	        & 1,445      &	156     &  	    1,273    &	    153\\
    31  &    North Yorkshire	    & 1,330      &	93      &   	990	     &       88\\
    32  &    Suffolk	            & 1,324      &	155     &  	    1,084    &	    140\\
    33  &    Cheshire	            & 1,254      &	125     &  	    1,061    &	    114\\
    34  &    Hertfordshire	        & 1,221      &	132     &  	    794	     &       100\\
    35  &    Gloucestershire	    & 1,163      &	144     &  	    656	     &       89\\
    36  &    Humberside	            & 1,160      &	120     &  	    971	     &       114\\
    37  &    Cambridgeshire	        & 995       &   71      &   	687	     &       64\\
    38  &    Derbyshire	            & 989       &   145     &  	    627	     &       128\\
    39  &    North Wales         	& 778       &   120     &  	    672	     &       110\\
    40  &    Cleveland	            & 732       &   58      &   	395	     &       53\\
    41  &    Cumbria	            & 615       &   107     &  	    526	     &       100\\
    42  &    Warwickshire	        & 499       &   48      &   	437	     &       41\\
    43  &    Gwent	                & 483       &   55      &   	323	     &       53\\
    44  &    Durham	                & 408       &   33      &   	263	     &       26\\
    45  &    Dyfed-Powys	        & 318       &   50      &   	304	     &       49\\
    \bottomrule
  \end{tabular}
   \caption{UK Policing regions ranked by volume of Malicious Scanning activity originating from them, within the study period Dec '20 to Dec '21.}
\end{table*}

\section{Questions, implications and opportunities}
In undertaking this research process we set out to examine if \textemdash through analysis of the volume, type and locations of detectable and therefore identifiable malicious scanning activities \textemdash it is possible to develop an evidence-base upon which one may be able to develop a multi-layered cyber threat intelligence model for the UK. In considering this we note that the data gathered evidences that it is possible to gather quantitative data regarding scanning activity from specific geographic regions, be these occurrences, cities, policing regions. Using this data it is possible to identify potentially malicious scanning activity that, in some instances, is indicative of known forms of detectable and therefore potentially preventable Malware. This tell us that it is possible to collect data which can be used to develop a UK focused threat intelligence model. For this model to be multi-layered then it should cross verticals, and be valuable to technical and non-technical stakeholders. In considering these requirements we note that the data gathered is not restricted to a single domain and in this respect fulfills part of our objectives. In terms of being valuable to technical and non-technical stakeholders, we must further consider if the insights gained from data analysis may suggest specific actions that could create value for technical and non-technical stakeholders in different verticals.\\ 
\\
The data examined suggests that the top 5 most frequently tagged malware scanning activities accounts for 46.34\% of the total volume of malware tagged in the period of study. If the data gathered is indicative of typical trends over a much longer period, then we argue that a campaign of activity directed towards the prevention, detection and response to emph{Mirai}, Web Crawlers, Telnet Brute Forcers, IoT brute forcers, and emph{Eternalblue} could improve the cyber-defences of UK citizens, charities and businesses. In arguing this we fully recognise that some of the scanning activities are much more challenging to address than others. Nonetheless, we propose that focusing part of the UK's civil cyber defence efforts on improving the understanding of technical and non-technical stakeholders with regards how to prevent, detect and respond to these threats could positively improve UK cyber defence. In suggesting this we also propose that a practical pilot study could be designed which tests the efficacy of taking such a targeted approach.\\    
\\
In the study we noted that IP addresses can be resolved back to the ASN controllers. Some of these ASN controllers are also Internet Service Providers who lease IPs to domestic users, businesses and charities. An interesting finding of this study is that annually somewhere between 0\% and 0.04\% of almost all ASN controllers' IP addresses are associated with performing malicious scanning activity. This is a comparatively small percentage of their overall pool and given their responsibilities towards securing their own networks, the question arises could ASN controllers do more to alert users that their IP address is associated with malicious scanning activity, so as to support broader campaign of activity to improve UK cyber defence?\\
\\
Building on the above it was noted that, through working with ASN controllers it may be possible to resolve malicious scanning activities back to companies, individuals and potentially even devices in networks.  To achieve this would require collaboration between ISPs and a publicly trusted 3rd party. If this were undertaken as a public service in order to provide  private citizens, SMEs and charities with a publicly accessible tool through which to check their own IP addresses for malicious activity, then this could offer a means to create further value for both technical and non-technical stakeholders. In suggesting this we also note that if such a collaboration were undertaken as a security or policing endeavour then we speculate that this could invoke similar negative reactions as took place with the \emph{Investigatory Powers Act (2016)} \cite{InvestigatoryPowersAct2016}, thus caution and an exploratory pilot study is advised to test the potential of this suggestion.\\
\\
The final question that remain to be addressed is that of how the data collected may inform a WEP-style language that could be useful to technical and non-technical stakeholders across different verticals. Having analysed our data set, an obvious shortcoming is that because the data is comprised of detectable and therefore potentially preventable threats, it only covers part of the spectrum of threats faced by UK cyber defenders. Thus it does not provide an adequate basis to inform the complete requirements of a WEP-style language.  Further, having considered what the data does and does not tell us about the nature of cyber threats, it is questionable what purpose a WEP-style language would serve since from a technical perspective the utility of attribution is limited. In balance, we must also consider that while cyber incidents directly impact technical systems, they also indirectly, though often purposefully, impact  non-technical domains.  Thus we argue that that the value of a WEP-style language lies not in its ability to attribute cyber incidents to specific threat actors but rather in its potential to classify the character and severity of a threat in terms that enable precise communication between technical and non-technical stakeholders when attempting to prevent, detect or respond to specific cyber incidents. This may therefore enable technical and on-technical stakeholders to calibrate their efforts. To this end we return to the categories of Cyber Criminal, State and State-Sponsored Actors, Terrorists, Hacktivists and Script Kiddies \cite{NationalCyberSecurity2016} and propose that the relationship between political motivation and technical capability form an \(X\), \(Y\) axis within which each of the five given categories could be plotted. Thus the data provides a starting point for further research in this domain.

\section{Limitations}
This study would not be complete without some discussion of its limitations. A key limitation is that the data presented is collected via a single source \textemdash \emph{Greynoise} \textemdash with further data collection and contextualization from the other cited sources. While \emph{Greynoise} has a formidable collection system spanning many networks and countries, we have to recognise the limitations of a single source.  Another potential limitation is that our research is focused entirely on the UK. Thus we have not considered if the scanning activities we have observed in relation to the UK are typical or anomalous when compared to other countries. These limitations of data-source and geographic focus also present opportunities for further research and thus it may be possible to develop more informed insights.\\ 
\\
Another note worthy limitation of our work is that the period of collection, though long enough for our initial study, is in comparative terms relatively short. It has therefore been a challenge to develop insights that would be indicative of trends in scanning behaviour. Therefore in future work we would aspire to work over a longer period of time, reporting results periodically. With these shortcomings in mind, we believe that our findings though modest establish a basis for future research that would enable us to more completely explore the questions we outlined in the background and motivation section of this text.

\section{Conclusion}
Through this study we have developed an evidence-based approach to UK cyber defence with a specific focus on addressing the types of cyber security threats faced most frequently by UK citizens. We have achieved this through a focus on detectable and therefore potentially preventable forms of malicious scanning activity, many of which are symptomatic of devices infected with malware. The findings of the study highlight those types of scanning activity which are most prevalent in the UK and provide suggestions for targeted strategies that may be initiated in order to reduce these activities. In doing this we also identify major stakeholders who could potentially work together for the public good and help reduce the threats identified herein. We believe their is sufficient evidence gathered to warrant the development of pilot schemes to explore practical approaches to supporting domestic internet users, SMEs and charities in better defending themselves against malicious internet scanning activities that target or infect their systems.\\
\\
%%\bibliographystyle{ACM-Reference-Format}
%%\bibliography{references.bib}
%%\end{document}
%%\endinput

\section*{Acknowledgments}
This research was undertaken while completing MSc studies in Software and Systems Security at the University of Oxford, UK. With gratitude, I acknowledge the advice and encouragement I received from Professor Andrew Simpson while progressing this research.

%Bibliography
\bibliographystyle{acm}  
\bibliography{references.bib}  
\end{document}